\documentclass[twocolumn,twoside]{osajnl}
\usepackage{amsmath,amssymb,amsfonts,bbm,graphicx,color}
\journal{josab}
\newcommand{\ket}[1]{\vert #1 \rangle}
\newcommand{\bra}[1]{\langle #1 \vert}

\def\ID{{\scriptscriptstyle I\!D}}
\def\QQ{{\scriptscriptstyle Q}}
\def\AC{{\scriptscriptstyle A\!C}}

\DeclareMathOperator{\tr}{tr}
\author[1]{Hamza Adnane}
\affil[1]{Laboratoire de Physique Th\'eorique, Facult\'e des Sciences Exactes,\\
	Universit\'e de Bejaia, 06000 Bejaia, Algeria}
	\author[2]{Berihu Teklu}
\affil[2]{Center for Cyber-Physical Systems, Khalifa University, Abu Dhabi, UAE}
	\author[3]{Matteo G. A. Paris}
\affil[3]{Quantum Technology Lab, Dipartimento di Fisica 'Aldo Pontremoli', Universit\`a degli Studi di Milano, I-20133 Milano, Italy}
\title{Quantum phase communication channels assisted by non-deterministic 
noiseless amplifiers}
\dates{Compiled \today}
\ociscodes{(270.5565)  Quantum communications, (270.5585)  
Quantum information and processing, (140.4480) Optical amplifiers,
(120.5050)  Phase measurement, (270.5570) Quantum detectors}
\doi{}
\begin{abstract}
We address quantum $M$-ary phase-shift keyed (PSK) communication 
channels in the presence of phase diffusion, and analyze the use 
of probabilistic noiseless linear amplifiers (NLA) to enhance 
performance of coherent signals. We consider both static and 
dynamical phase diffusion and assess the performances of the 
channel for ideal and realistic phase receivers. Our results show 
that NLA employed at the stage of signal preparations is a useful 
resource, especially in the regime of weak signals. We also discuss 
the interplay between the use of NLA, and the memory effects 
occurring with dynamical noise, in determining the capacity of the 
channel. 
\end{abstract}
\setboolean{displaycopyright}{true}
\begin{document}
\maketitle
\thispagestyle{fancy}
\section{Introduction}
Quantum communication channels based on continuous variables (CV) 
have attracted an increasing interest in the recent years, due to their 
robustness against noise~\cite{Holevo01}.  For lossless CV channels, 
the capacity at fixed energy is maximised by thermal encoding of information 
onto Fock states\cite{Yuen}. On the other hand, when propagation and 
detection are affected by loss and/or noise, alternative strategies, where 
information is encoded onto either the phase or the amplitude of coherent 
signals, have proven effective \cite{Giovannetti,gio04}.
\par
In a phase modulation scheme, where the information is encoded in the 
phase of a quantum seed signal~\cite{Nair,D'Ariano}, the most detrimental 
noise is phase diffusion~\cite{Olivares,Genoni}. In 
particular,  when the seed state is coherent, it has been shown that 
time-independent Markovian noise is detrimental to information transfer 
and may undermine the overall performance of the channel~\cite{Teklu,Trapani}. 
However, in quantum optical communications, the Markovian hypothesis 
may be violated by the spectral structure of the environment leading to 
non-Markovian damping or diffusion channels~\cite{Trapani,Vasile}. 
Thereby, reservoir engineering may lead to substantial improvements 
in optical communication channels, by properly handling the unavoidable 
interaction with the environment during propagation.
\par
More generally, in order to reduce the detrimental effects of the 
noise, and the corresponding loss of information, different types of 
amplification processes may be employed. None of them, however, is 
expected to restore ideal conditions, due to the inherent quantum limits 
of amplification. In particular, the noise figure $R$ 
(i.e. the ratio between the input and the output, signal-to-noise ratios) 
of linear quantum amplifiers is bounded by $R>2$, leading to the well 
known $3$dB standard quantum noise limit \cite{Caves82}. In this 
paper, we exploit the possibility of overcoming a possible this fundamental 
quantum limitation is by means of probabilistic amplification based on 
conditional dynamics in quantum phase communication channels. Furthermore, 
we investigate the use of probabilistic (and noiseless) linear amplifier (NLA) at the stage 
of signal preparation to improve the performances of noisy 
phase channels based on coherent signals.
\par
We consider a protocol where information is encoded in the phase-shift
of a seed state, which is then transferred to a receiver station along 
a transmission line where phase diffusion (static noise) or phase 
fluctuations (dynamical noise) may occur. We consider 
static phase noise induced by a Markovian environment as well as 
a dynamical noise leading to non-Markovian evolution. We evaluate 
the mutual information for NL-amplified coherent states for both 
ideal and realistic phase receivers at the detection stage. 
In practice, the successfully (heralded) amplified 
states serve as careers for phase information transmission processes, 
and undergo the whole process of encoding-transfer along 
the noisy channel-decoding, whereas when the amplification fails, the 
sender abstains from imprinting letters in the distorted states and 
discard them instead. We then compare their performances and also compare those
phase channels to noisy amplitude-based ones, where information is encoded onto the 
amplitude of coherent states. Finally, we discuss the interplay 
between the use of NLA and the memory effects occurring with dynamical
noise in determining the capacity of the channel. 
\par
Among the different possible implementations of NLA schemes, we 
focus attention on a feasible one \cite{Jaromir} which, in turn, 
has been experimentally achieved with current technology~\cite{Zavatta},
at least for a given values of the gain. Our results are therefore
of practical interest and may be experimentally verified. 
\par
The paper is organized as follows. In sec~\ref{s:sec2}, we review 
the model of the probabilistic linear amplifier considered 
in our work and address its action on coherent states. In section~\ref{s:sec3}, 
we describe in some details of the dephasing dynamics. We then review, in sec~\ref{s:sec4} 
the main steps of the encoding-decoding strategy in the quantum phase channel. 
Sec~\ref{s:sec5} is devoted to illustrate our results for both static and
 dynamical  phase noise. Finally, sec~\ref{s:sec6} closes the paper 
 with some concluding remarks.
\section{Model of the noiseless linear amplifier}\label{s:sec2}
An ideal, noiseless deterministic amplification of a quantum state is 
inherently forbidden by the unitarity and linearity of quantum evolution. 
In fact, so as to reinstate the uncertainty principle, an unavoidable 
noise has to be introduced. A quantum-noise limit has been drawn up for 
both the two versions of linear deterministic amplifiers: phase-sensitive 
and phase-insensitive~\cite{Caves82}. Recently, much attention 
has been devoted to a new generation of linear amplifiers whose action 
differs from the conventional devices: probabilistic noiseless amplifiers~\cite{Ralph09}. 
In turn, the non-deterministic nature of those devices enables to elude the 
theoretical limitation prohibiting noiseless amplification. Hereafter, several theoretical 
schemes and implementation of the NLA have been proposed~\cite{Usaga,Ferreyrol,McMahon,Pandey}. 
Here, we consider the theoretical model suggested in \cite{Jaromir} and experimentally demonstrated in 
\cite{Zavatta} for a given value of the gain.  
\par
Let us now describe the noiseless amplification in the Schr\"odinger picture, 
where it may be perceived as the operation that takes a coherent state
 $\ket{\alpha}$ to its approximated amplified version $\ket{g\alpha}$ 
 with a rate of success $p_{s}$ that depends on the mean energy of the 
 input state. A probabilistic noiseless amplification may thus be 
 described by the non-unitary operator 
\begin{equation}\label{key}
{\mathcal{M}}(g)=g^{{a^\dag a}}\,,
\end{equation}
$a$ being the field mode operator. The instance we are interested is $g>1$, 
where $g$ refers to the {\em gain} of the amplifier. The action of such 
operator on Fock states $\ket{n}$ consists to assign a factor $g^{n}$ 
to their amplitudes. The operator ${\mathcal{M}}(g)$ is unbounded, and 
the physical consequence of this mathematical issue is that a noiseless 
amplification may only be approximately achieved with a finite 
success probability. In particular, it may be implemented by truncating 
the expansion of $g^{{a^\dag a}}$ to a given order in the number operator,
that we refer to as the truncation order. This approximation is well 
justified for the class of weak signals we are going to consider in 
the following. In particular, let us consider the case where the truncation 
order is set to one. A good figure of merit to assess the performance of
the amplification process is the so-called effective gain, defined as 
the ratio of the amplitude of the amplified state $${\varrho}_{\text{a}}=\frac{\mathcal{M}(g)\ket{\alpha}\bra{\alpha}\mathcal{M}^{\dagger}(g)}{\tr\left[ \mathcal{M}(g)\ket{\alpha}\bra{\alpha}\mathcal{M}^{\dagger}(g)\right] }\,,$$ to that of 
the input coherent state, i.e.
\begin{equation}\label{}
g_{\text{eff}}=\frac{1}{\alpha}\tr\left[{a}\,{\varrho}_{\text{a}} \right]. 
\end{equation}
Fundamental constraints to noiseless amplification require the fidelity of the output state to the ideally amplified coherent state to approach unit value and the effective and nominal gains to coincide $g_{\text{eff}}=g$ in the limit of vanishing input energies. Upon truncating the Taylor series expansion of the NLA operator in its first order and fulfilling the previously raised constraints, we derive the expression of the approximate NLA operator \cite{Zavatta}
\begin{equation}\label{amplification op}
\mathcal{M}(g)=1+\left(g-1\right){a^\dag a}. 
\end{equation}
In the following, we refer to the state resulting from the action of the 
approximate NLA in Eq. (\ref{amplification op}) on coherent input as 
(AC) $\ket{\alpha}_{\text{a}}$. Its density matrix elements in the Fock 
basis may be expressed as follows
\begin{equation}\label{alpha amplified}
\varrho_{ n,m}=\frac{e^{-\bar{n}}}{\mathcal{A}}
\frac{\bar{n}^{\frac{n+m}{2}}}{\sqrt{n!\, m!}}
\big[1+(g-1)n\big]
\big[1+(g-1)m\big]\,,
\end{equation}
where $\bar{n}$ denotes the average photons number and $\mathcal{A}^{-1}$ a normalisation constant given by
\begin{equation}\label{}
\mathcal{A}=1+\left( g^{2}-1\right) \bar{n}+\left( g-1\right) ^{2}\bar{n}^{2}\,,
\end{equation}
As it is apparent from Eq. (\ref{amplification op}), the NLA with 
gain $g=2$ reduces to photon addition and photon subtraction 
\cite{Oli2005} performed in a sequence, a procedure which is
experimentally available~\cite{Zavatta} with current
technology. It is worth noting that the model of NLA 
considered throughout the paper presents several advantages 
compared to other physical schemes, as demonstrated by its 
performances quantified by the effective gain and 
fidelity~\cite{Zavatta}. Besides the realistic implementations 
mentioned so far, more abstract schemes for noiseless amplification have
been recently proposed~\cite{Jarumir,Walk,Chrzanowski}. Their 
operating principle is based on a postselection of classical 
data collected from heterodyne detection that emulate a noiseless 
amplification. Despite being experimentally friendly, as only 
feasible Gaussian operations are required, the emulated NLA is 
not suited for our protocol due to the restrictive need of 
being directly followed by a heterodyne detection, thereby 
embedded within the detection stage.
\section{Static and dynamical phase diffusion}\label{s:sec3}
This section is devoted to model the dynamics induced by a phase diffusive 
classical environment on a continuous variable system. In particular, we show 
that treating the environment as a classical stochastic field (CSF) provides 
an effective description of the dynamics, able to describe rich phenomenology 
that canonical Master Equations approach, usually derived from too restrictive 
approximations, may not capture. 
\par 
Let us consider a single bosonic mode interacting with a classical stochastic 
field (CSF). The dynamics of the system is governed by the total Hamiltonian, i.e. 
the sum of the free and interaction Hamiltonians given respectively by
\begin{equation}\label{}
{H}_{0}=\hbar\omega_{0}{a}^{\dagger}{a} 
\end{equation}
\begin{equation}\label{Hamiltonian}
{H}_{i}=\hbar{a}^{\dagger}{a} \left[F(t)e^{-i\omega t}+F^{*}(t)e^{i\omega t}\right] 
\end{equation}
where $\omega_{0}$ is the proper frequency of the oscillator and $F(t)e^{-i\omega}$ an external fluctuating field with a complex amplitude that oscillates in time with a central frequency $\omega$. $F(t)$ describes the realizations of a stochastic 
process with zero mean, accounting for the noise induced by the surrounding 
degrees of freedom and $F(t)^{*}$ denotes its complex conjugate. 
As the interaction and free Hamiltonians commute, the Hamiltonian operator 
in the interaction picture reads (in units of $\hbar$) 
\begin{equation}\label{HamiltonianI}
{H}={a}^{\dagger}{a} \left[F(t)e^{-i\omega t}+F^{*}(t)e^{i\omega t}\right] 
\end{equation}
where the stochastic field $F(t)$ assume a dimension of frequency.
The Hamiltonian being self-commuting at different times, i.e. $\left[{H}(t),{H}(t^{'}) \right]=0 $, the evaluation of the evolution operator expression is made straightforward. As it appears, the evolution operator coincides with a unitary phase-shift :
\begin{equation}\label{evolutionop}
{U}(t)=e^{i\phi(t){a}^{\dagger}{a} }
\end{equation}
where $\phi(t)=\int_{0}^{t}ds\left[ F(s)e^{-i\omega s}+F(s)^{*}e^{i\omega s}\right] $ accounts for the phase-shift performed on the system and depends essentially on the (CSF). After expanding the initial state in a Fock basis, the evolved density matrix at time $t$ reads 
\begin{align}
\label{rhoevolved}
{\varrho}(t)=&\left\langle {U}(t){\varrho}(0){U}^{\dagger}(t)\right\rangle_{F}  \\
=&\sum_{n,m}\left\langle e^{-i\phi(t)(n-m)}\right\rangle _{F}\varrho_{nm}\ket{n}\bra{m}, 
\notag 
\end{align}
where $\left\langle. \right\rangle_{F} $ represents the average over all possible realisations of the stochastic process. From now on, we consider Gaussian stochastic processes that are fully characterized by their two first order statistics, that is, the mean $\mu(t)=\left\langle F(t) \right\rangle _{F}$ and the autocorrelation function $K(t,t^{'})=\left\langle F(t)F(t^{\prime}) \right\rangle _{F}$. In particular, we will focus on stochastic fields with zero mean and autocorrelation matrix assuming a diagonal form
\begin{equation}\label{autocorrelation}
\left\langle \mathcal{R}\left[ F(t)\right] \mathcal{R}\left[ F(t^{\prime})\right] \right\rangle _{F}=\left\langle \mathcal{I}\left[ F(t)\right] \mathcal{I}\left[ F(t^{\prime})\right] \right\rangle _{F}=K(t,t^{'})
\end{equation}
\begin{equation}\label{autocorrelation2}
\left\langle \mathcal{R}\left[ F(t)\right] \mathcal{I}\left[ F(t^{\prime})\right] \right\rangle _{F}=\left\langle \mathcal{I}\left[ F(t)\right] \mathcal{R}\left[ F(t^{\prime})\right] \right\rangle _{F}=0,
\end{equation}
where $\mathcal{R}\left[ F(t)\right] $ and $\mathcal{I}\left[ F(t)\right] $ denote respectively the real and the imaginary parts of the stochastic field. The rearrangement of the phase $\phi(t)$ in terms of two distinct contributions coming from the real and imaginary parts of the (CSF) simplifies remarkably the evaluation of the evolved density matrix. The average in Eq~(\ref{rhoevolved}) reduces to the evaluation of the joint characteristic function of two Gaussian variables 
\begin{equation}\label{}
\left\langle e^{-i\phi(t)(n-m)}\right\rangle _{F}=e^{-\frac{1}{2}(n-m)^{2}\sigma(t)},
\end{equation}
with $\sigma(t)$ being a function that depends on the kernel of the stochastic process
\begin{equation}\label{sigma}
\sigma(t)=\int_{0}^{t}\int_{0}^{t}dsds^{\prime}cos\left[\omega(s-s^{\prime})\right] K(s,s^{\prime}).
\end{equation}
The evolved density matrix of Eq~(\ref{rhoevolved}) then simplifies to
\begin{align}
\label{rhoevolved1}
{\varrho}(t)=\sum_{n,m} e^{-\frac{1}{2}(n-m)^{2}\sigma(t)}\varrho_{n,m}\ket{n}\bra{m}.
\end{align}
As it appears the diagonal elements $\varrho_{nn}$ are left unchanged under the phase diffusion, thus preserving the occupation probabilities, while the off-diagonal matrix elements vanish exponentially.
When $\sigma(t)$ displays a linear behaviour in time, the Gaussian process is 
said to be {\em static}. In that case, one recovers the solution of the Master 
Equation that governs the evolution of a system undergoing Markovian 
phase diffusion noise. We defer a detailed discussion for the following parts 
where a specific model of Gaussian stochastic process (the power-law process) 
will be analyzed.   
\section{Quantum phase-shift keyed communication}\label{s:sec4}
In phase-modulation-based communication channels $M$ symbols selected from 
a given ensemble are encoded using $M$ uniformly spaced phase shifts $\phi_{l}$ 
ranging from $0$ to $2\pi$. The encoding procedure is carried out by a 
phase-shift operation ${U}(\phi_{l})$ on a seed single-mode state 
${\varrho}_{0}$ that 
yields the deterministic state ${\varrho}_{l}={U}(\phi_{l}){\varrho}_{0}{U}^{\dagger}(\phi_{l})$. Next, the signal ${\varrho}_{l}$ is sent through a transmission 
 line to a receiving station where a phase measurement, followed by a suitable 
 inference strategy are performed so as to extract the information. A straightforward strategy consists on dividing the ensemble of possible outcomes to $M$ intervals 
 $$\Sigma_{l}=\left[ \phi_{l}-\frac{\Delta}{2},\phi_{m}+\frac{\Delta}{2}\right)\,, 
 $$ of width $\Delta=2\pi/M$ and associate each measured phase that falls into 
 $\Sigma_{l}$  with the corresponding symbol that the phase-shift $\phi_{l}$ accounts for. It's evident that the intervals $\Sigma_{l}$ sum up to $\left[ 0,2\pi\right)$. 
We point out that, naturally, one may opt for a different inference strategy where a non-uniform width of the intervals is chosen. As a symmetric choice is generally optimal, we adopt the previously-presented scheme that may be summarized in the following terms: for each outcome $\phi$ of the phase measurement, if $\phi\in\Sigma_{l}$, then $\phi\mapsto\phi_{l}$.
The statistics of a phase measurement outcomes are described by a positive 
operator-valued measure (POVM) $\left\lbrace \pi(\theta)\right\rbrace $ 
where $\theta\in\left[ 0,2\pi\right)$. The recipe that provides the 
probabilities of the receiver's outcomes $\Pi_{l}$ reads as~\cite{Leonhardt}
\begin{equation}\label{Pil}
{\Pi}_{l}=\int_{\Sigma_{l}}\pi(\theta)d\theta.
\end{equation}  
A covariant phase measurement can always be described by a POVM $\left\lbrace \pi(\theta) \right\rbrace $ that assumes the following form:
\begin{equation}\label{pi}
{\pi}(\theta)=\frac{1}{2\pi}\sum_{n,m=0}^{\infty}\,
A_{n,m}\,e^{-i(n-m)\theta}\ket{n}\bra{m},
\end{equation}
where $A_{n,m}=2\pi\bra{m}{\pi}(0)\ket{n}$ are the elements of a positive and Hermitian matrix $A$ set by the chosen phase measurement. We remark that the covariance of the phase measurement performed by the receiver is ensured by the covariance property 
of $\left\lbrace {\pi}(\theta)\right\rbrace $. This implies :
\begin{equation}\label{covariance}
{\Pi}_{l}(\theta)={U}(\phi_{l}){\Pi}(0){U}^{\dagger}(\phi_{l}).
\end{equation}
Starting from Eqs (\ref{pi}) and (\ref{Pil}), we arrive to the compact formula 
\begin{equation}\label{Pitheta}
{\Pi}_{l}(\theta)=\sum_{n,m=0}^{\infty}A_{n,m}f_{n-m}(l)\ket{n}\bra{m}
\end{equation}
where $f_{d}(l)=\frac{1}{2\pi}\int_{\Sigma_{l}}e^{-id\theta}d\theta$  refers to as the resolution function. Since the phase-shifts are equidistant ($\phi_{l}=2\pi l/M$) and the range of possible outcomes is uniform, the resolution function assumes 
the following form
\begin{equation}\label{resolutionfunc}
f_{d}(l)=\frac{e^{-\frac{2\pi ld}{M}}}{\pi d}\sin{\frac{\pi d}{M}}.
\end{equation}
\par
In order to quantify the performance of the phase-shift keyed communication 
channel, we employ the mutual information (MI) between output and input, as 
a suitable figure of merit that measures the amount of information transferred 
along the transmission line at each use of the channel. MI is given by: 
$I=S(O)-S(O| I)$, $S(O)=\sum_{l=0}^{M-1}p^{\prime}(l)
\log{\left( p^{\prime}(l) \right) }$ being the total information 
available at the receiver (output) and $S(O| I)=
\sum_{k=0}^{M-1}p(k)S(O|k)=\sum_{l,k=0}^{M-1}p(k)p(l|k)\log{p(l|k)}$ 
the conditional information available at the output knowing which element 
($\phi_{k}$) from the input ensemble was sent averaged over the possible inputs. 
We clearly notice that the mutual information is determined by three quantities: the prior probability that a given symbol ($\phi_{k}$) carried by the state ${\varrho}_{k}$ was transmitted, the probability for the receiver to read out a given symbol ($\phi_{l}$), that reads $p^{\prime}(l)=\sum_{k=0}^{M-1}p(l|k)p(k)$ and finally, the conditional probability $p(l|k)$ for the receiver to measure a phase ($\phi_{l}$) given that the input symbol encoded in ($\phi_{k}$) was transmitted. 
The classical channel capacity, namely the maximum information reliably transferred through the transmission line per use is given by the maximum over the prior probability $p(k)$ of the mutual information. Throughout this paper, we assume a uniform prior. 
In other words, the states ${\varrho}_{k}$ are transmitted according to the same probability i.e. ($p(\phi_{k})\equiv p(k)=1/M$). Hence, the evaluation of the mutual information depends mainly on the conditional probability $p(j|k)$ and its expression reads 
\begin{equation}\label{MI}
I=\frac{1}{M}\sum_{l,k=0}^{M-1}p(l|k)\log{\frac{Mp(l|k)}{\sum_{k=0}^{M-1}p(l|k)}}.
\end{equation}
The conditional probability $p(l|k)$ is thought as the probability that a measurement outcome belongs to the phase interval $\Sigma_{l}$ when the state ${\varrho}_{k}$ has been actually transmitted along the channel. Owing to the covariance property of the POVM ${{\Pi}_{l}}$, its expression can be written as
\begin{align}
\label{conditional}
p(l|k)=&\tr{\left[{\varrho}_{k}{\Pi}_{l}\right]} =\tr{\left[{\varrho}_{0}{\Pi}_{l-k} \right] }  \\
	=&\sum_{n,m=0}^{\infty}A_{n,m}f_{n-m}(l-k)\varrho_{n,m} .\notag 
\end{align}
Besides, making use of the symmetries of the resolution function $f_{-d}(-l)=f_{d}(l)$, and upon introducing the positive index $s=\left|l-k \right|$ ranging from 0 to $M-1$, the mutual information (\ref{MI}) simplifies to
\begin{equation}\label{MIf}
I\equiv I(M,\bar{n})=\log{M}+\sum_{s=0}^{M-1}q(s)\log{q(s)},
\end{equation}  
with $\bar{n}$ being the average photons number of the input state and $q(s)=\sum_{n,m=0}^{\infty}A_{n,m}f_{n-m}(s)\varrho_{n,m}$ a function that can be perceived as the probability that a phase difference of $\phi_{s}=2\pi s/M$ between the input and output signal is registered independently of the phase imprinted at the sending 
station. 
\par
Without loss of generality we may assume real matrix elements ${\varrho}_{n,m}$. 
It follows that the function $q(s)$ becomes
\begin{equation}\label{qs}
q(s)=\frac{1}{M}\left[1+\frac{2M}{\pi d} \sum_{n=0}^{\infty}
\sum_{d=1}^{\infty}A_{n,n+d}\varrho_{n+d,n}\cos{\frac{2\pi ds}{M}}\sin{\frac{\pi d}{M}}\right] 
\end{equation}
The mutual information of a phase-shift keyed (PSK) communication channel 
thus depends essentially on the seed state carrying the transferred information ($\varrho_{0}$), the intrinsic characteristics of the channel and the phase measurement performed at the stage of the receiver (defined through the matrix elements 
$A_{n,m}$). 
\par
In the present work, we analyze the performances of two specific phase measurements: the canonical phase measurement~\cite{Leonhardt,Paris99} and the angle margin of the Husimi Q-function~\cite{Noh,D'Ariano94}. We emphasize the feasibility of this latter via heterodyne or eight-port homodyne detection. Regarding the canonical measurement, the matrix elements are given by $A_{n,m}=1$ whereas for the phase-space-based measurement, we have $A_{n,m}=\frac{\Gamma\left[1+(n+m)/2 \right] }{(n!m!)^{1/2}}$, where $\Gamma\left[ x\right] $ is the Euler Gamma function. We also notice that due to its appealing properties, the canonical phase measurement is the optimal choice among the phase POVMs. As its physical implementation remains an open problem \cite{dp1,dp2}, we refer to 
as \textit{ideal} phase measurement. 
\section{Phase-shift keyed quantum communication channels in the presence of noise}\label{s:sec5}
Quantum communication channels are characterized by the seed state chosen at the stage of the preparation, the intrinsic properties of the channel, namely, the noise induced along the transmission line and the detection measurement performed by the receiver. This section is devoted to assess the performances of a quantum phase channel where the information is imprinted onto the phase of amplified coherent state (AC) in the presence of static and dynamic phase diffusion. The analysis will be drawn up either for ideal phase measurement and phase-space-based one, that we dub "Q-measurement" 
\subsection{Static phase diffusive channels}
Let us consider a travelling light beam in a static phase diffusive environment. Under the Born-Markov approximation, the evolution of the system is governed by the master equation:
\begin{equation}\label{}
\frac{d}{dt}{\varrho}(t)=\frac{\Gamma}{2}\mathcal{L}\left[{a}^{\dagger}{a} \right]{\varrho} ,
	\end{equation}
where $\mathcal{L}[{O}]{\varrho}=2{O}{\varrho}{O}^{\dagger}-{O}{O}^{\dagger}{\varrho}-{\varrho}{O}{O}^{\dagger}$ and $\Gamma$ denotes the phase noise factor. Given an initial seed ${\varrho}_{0}$, the time evolved density matrix is found to be
\begin{equation}\label{rhot}
{\varrho}(t)=\sum_{n,m=0}^{\infty}e^{-\frac{\tau}{2}(n-m)^2}\varrho_{n,m}\ket{n}\bra{m},
\end{equation}
with $\tau \equiv \Gamma t$ being the dephasing parameter. As mentioned previously, the phase diffusive noise doesn't affect the diagonal elements $\varrho_{n,n}$ (preserves the energy) while cancelling the coherences.
Let us now consider a seed signal prepared through the action of a non-deterministic noiseless amplifier on coherent states (AC). Throughout the paper, we consider coherent sates with zero initial phase. The density matrix elements of the seed signal are thus given by Eq (\ref{alpha amplified}). The transferred state along the noisy channel corresponds to the time evolved density matrix (\ref{rhot}) and its elements read: 
\begin{equation}\label{}
\varrho_{n,m}(t)=e^{-\frac{\tau}{2}(n-m)^2}\varrho_{n,m}.
\end{equation}

\begin{figure}[h!]
	\includegraphics[width=0.22\textwidth]{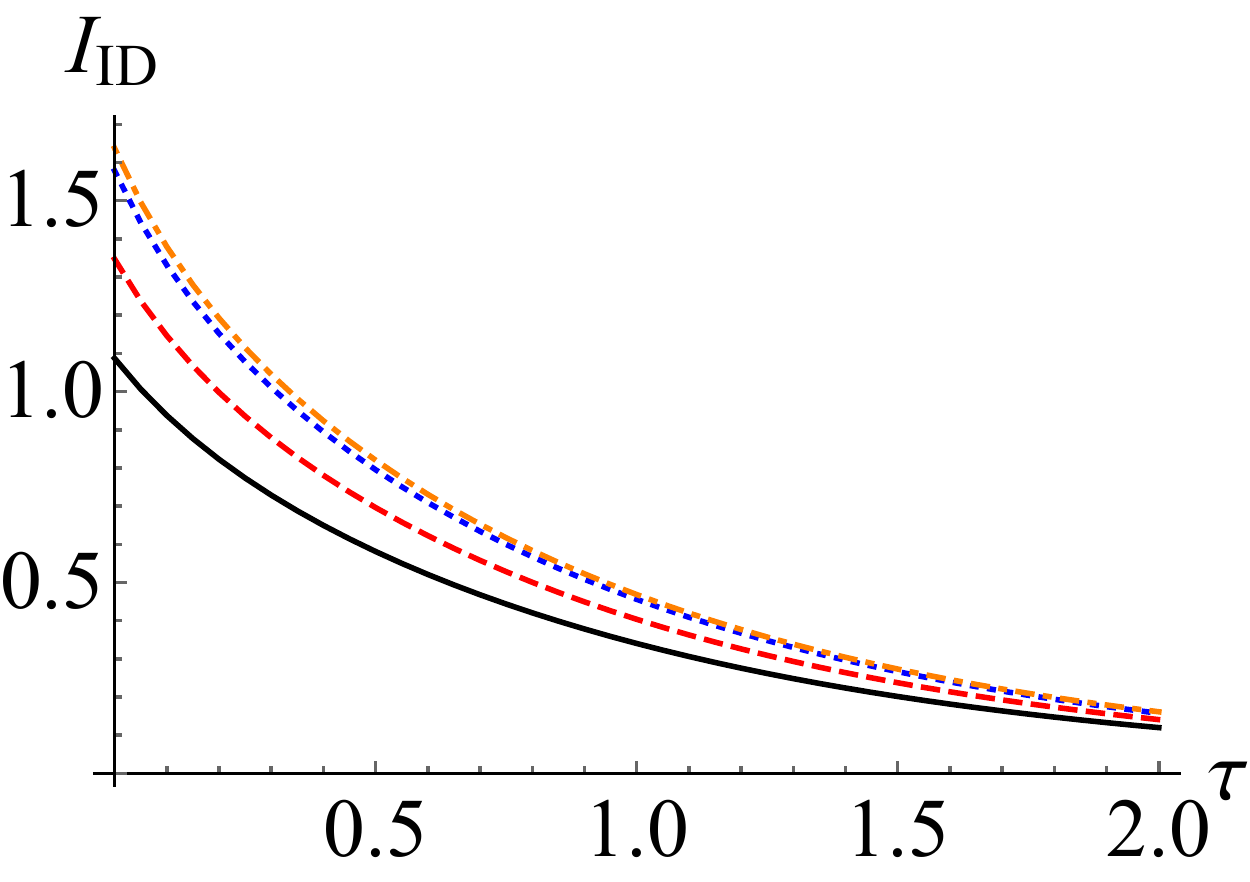}
	\includegraphics[width=0.22\textwidth]{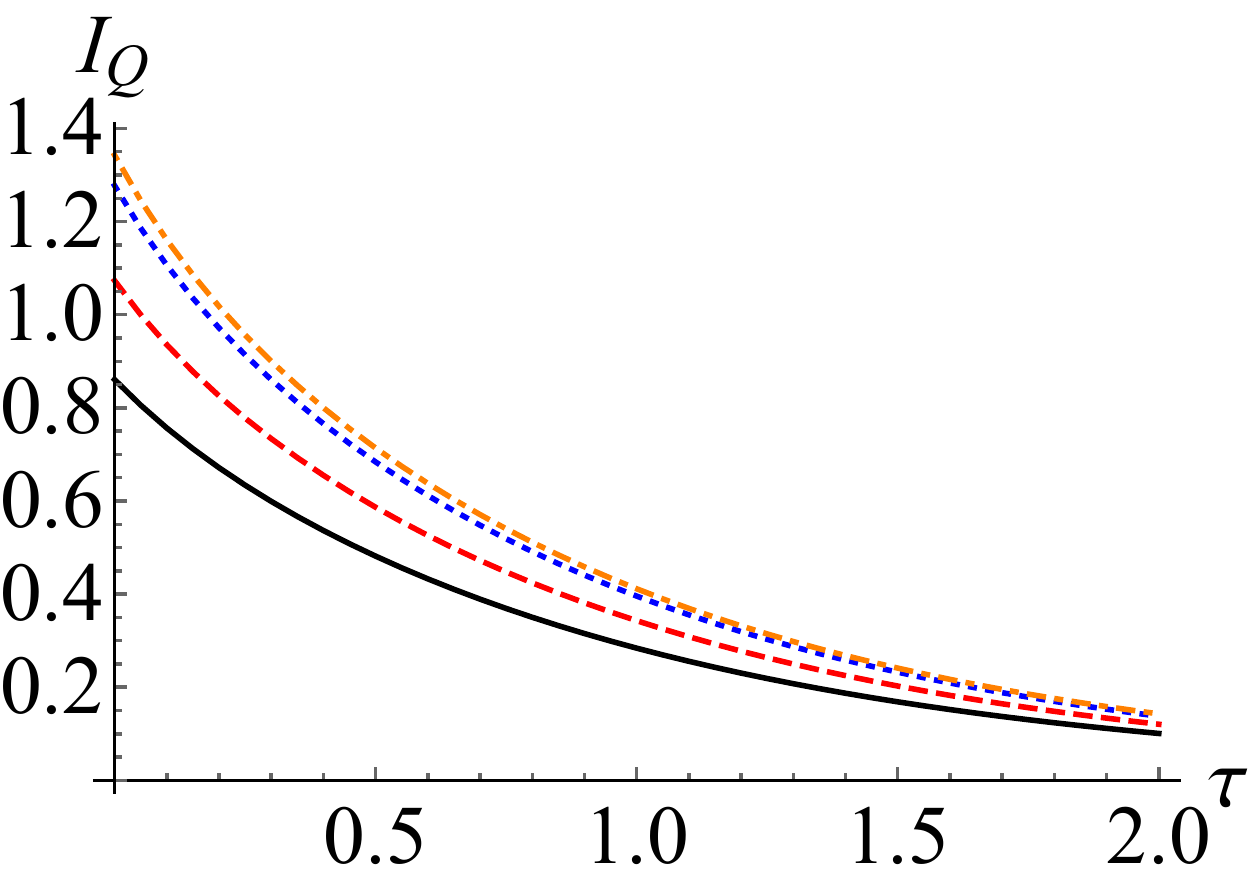}\\
	\includegraphics[width=0.22\textwidth]{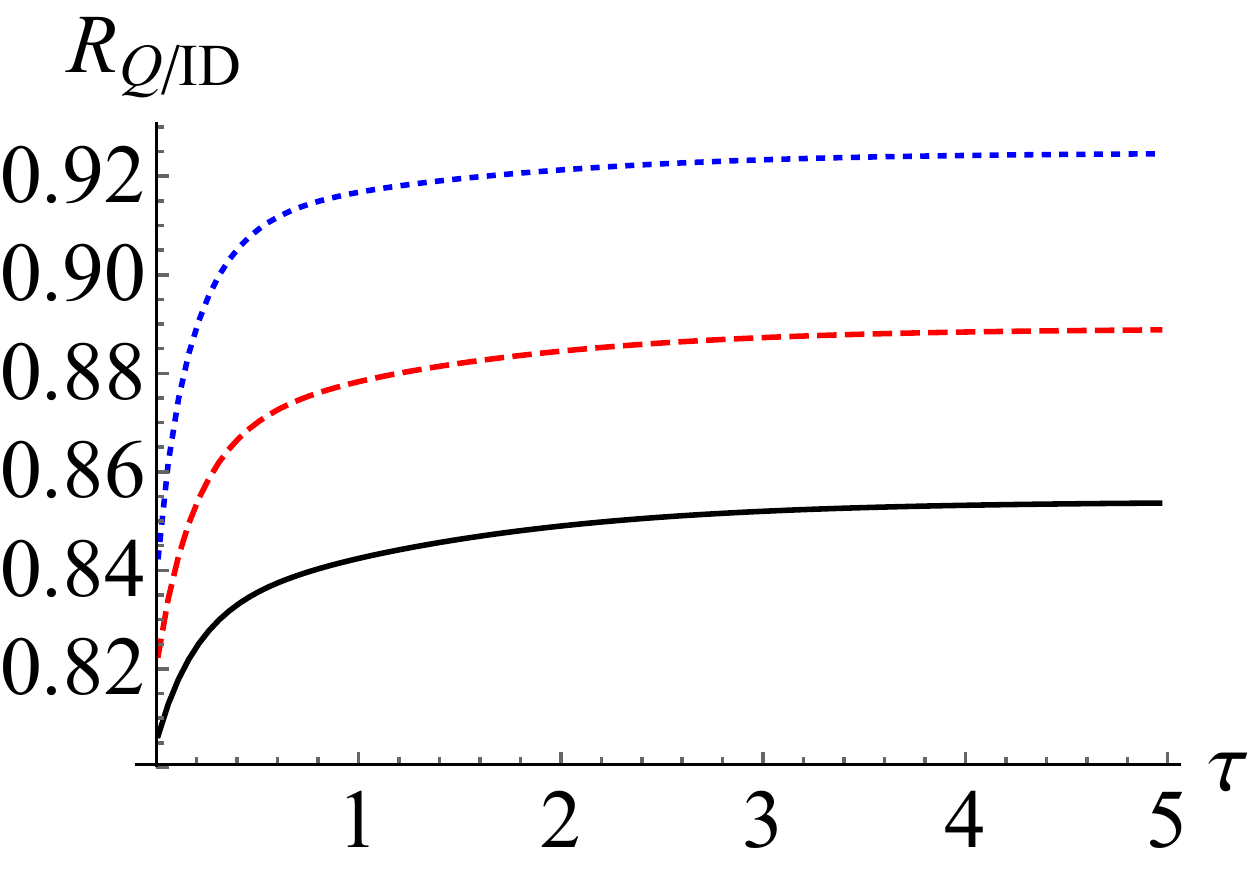}
	\includegraphics[width=0.25\textwidth]{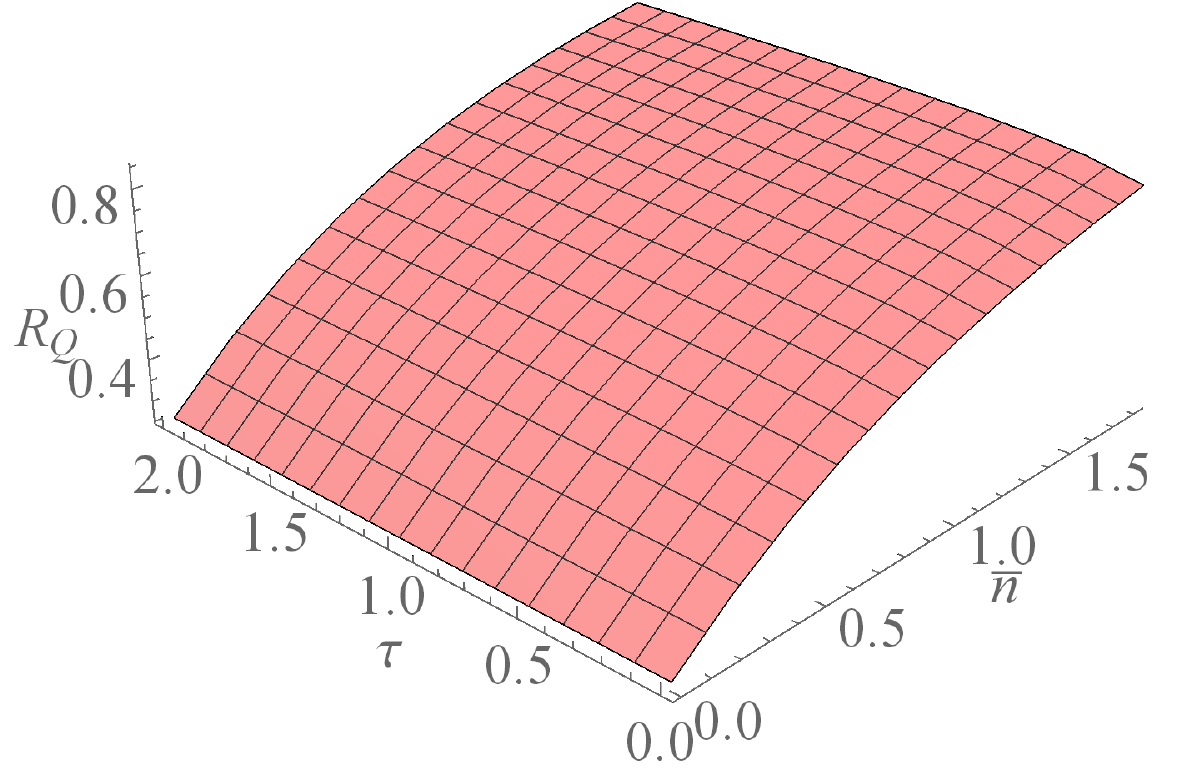}
	\caption{Color online) Upper panels: Performances of the ideal and Q receivers, namely, the mutual informations $I_{ID}$ (left) and Q $I_{Q}$ (right), in the presence of static noise as functions of the phase noise parameter $\tau$  for different seed signals: the solid black line represents the standard coherent state while the dashed red, dotted blue and dotdashed orange lines denote AC states generated respectively with the following NLA's configurations: $g=1.2,1.6,2$. The input energy of the primary coherent state $\bar{n}=1$.  Lower panel in left: The ratio $R_{Q/ID}=I_{Q}/I_{\ID}$ as a function of $\tau$ for different values of mean photons number: $\bar{n}=0.5$(solid black), $\bar{n}=1$(red dashed) and $\bar{n}=2$(blue dotted). Lower panel in right: 3D plot of the ratio $R_{\QQ}=I_{c}/I_{\AC}$ as a function of $\tau$ and $\bar{n}$. The symbols ensemble size is set to $M=20$ in all the plots.} \label{f:Fig1}
\end{figure}

Since the matrix elements of the received state are now identified, 
the evaluation of the mutual information now depends only on the kind
of phase measurement performed at the receiver's stage. Concerning 
ideal phase measurement, i.e. $A_{n,m}=1$, the probability 
$q(s)$ in the presence of static phase noise is given by
\begin{align}\label{qsI}
\begin{split}
q_{\ID}(s)=&\frac{1}{M}\Big[ 1+\frac{2M}{\pi d}\cos{\frac{2\pi ds}{M}}\sin{\frac{\pi d}{M}} \\
&\times \sum_{n=0}^{\infty}\sum_{d=1}^{\infty}\varrho_{n+d,n}e^{-\frac{\tau}{2}(n-m)^2}\Big] , 
\end{split}
\end{align}
whereas for Q-measurement the probability $q(s)$ reads:
\begin{align}\label{qsQ}
\begin{split}
q_{\QQ}(s)=&\frac{1}{M}\Big[1+\frac{2M}{\pi d}\cos{\frac{2\pi ds}{M}}\sin{\frac{\pi d}{M}} \\
&\times \sum_{n=0}^{\infty}\sum_{d=1}^{\infty}\frac{\Gamma\left[1+(n+m)/2 \right] }{(n!m!)^{1/2}}\varrho_{n+d,n}e^{-\frac{\tau}{2}(n-m)^2}\Big]  , 
\end{split}
\end{align}
The mutual informations $I_{\ID}$ and $I_{Q}$ of the two receivers are then evaluated by substituting expressions (\ref{qsI}) and (\ref{qsQ}) of $q_{\ID}(s)$ and $q_{\QQ}(s)$ in Eq (\ref{MIf}) respectively. In the upper panel of Fig~\ref{f:Fig1}, we report the mutual informations $I_{\ID}$ and $I_{\QQ}$ as functions of the dephasing parameter $\tau$ for different seed signals, corresponding to different values of  
the gain of the NLA. The mean photons number of the input coherent state at the 
stage of the seed preparation is set to $\bar{n}=1$, and the cardinality of 
the alphabet is fixed to $M=20$. We clearly notice the detrimental effects of the unavoidable phase noise on the amount of information transferred from the emitting station to the receiver. Nonetheless, the amplified coherent states tend to reduce the loss of information, yielding noticeable enhancement. As it is apparent from the plots, larger gains of the NLA better preserve the information flow either for an ideal receiver or a Q-measurement-based one. From now on, we will focus on the seed states 
prepared with an NLA calibrated such as its nominal gain is set to $g=2$. 
Our choice is motivated by the experimental feasibility of that particular 
amplifier. 
\par
As the mutual informations $I_{\ID}$ and $I_{\QQ}$ follow a similar qualitative behaviour, we report, in the same figure (bottom panel on the left), the behaviour 
of their ratio $R_{\QQ/\ID}=I_{\QQ}/I_{\ID}$ as a function of $\tau$, 
for different input energies $\bar{n}$. As expected, the ratio is not achieving 
unit value, thus confirming the optimality of the canonical phase measurement. 
On the other hand, as it may be noticed from the plots, when the energy of 
the signals increases, the performances of the Q receiver approach those 
of its ideal counterpart, especially for large values of the dephasing parameter.
In order to highlight the beneficial contribution of the NLA to enhance the mutual information, we use a 3D plot (bottom panel on the right) of the ratio $R_{\QQ}=I_{c}/I_{\AC}$ as a function of input energy $\bar{n}$ and the dephasing parameter $\tau$ for a number of symbols $M=20$. Here $I_{c}$ and $I_{\AC}$ account respectively for the mutual informations of a coherent seed signal and an amplified coherent seed (with $g=2$).  The 3D plot reveals that the mutual information obtained with 
the amplified coherent state (AC) surpasses that of the standard coherent 
seed signal. Furthermore, a monotonic increase of the ratio $R_{\QQ}$ with 
$\bar{n}$ for any fixed value of $\tau$ is noticed. Thus, the weaker is the 
input energy, the more substantial is the enhancement brought by the NLA. 
These results show the beneficial advantages of the NLA when used at the stage 
of the seed preparation, in particular in the regime of low energies.

We previously pointed out the optimality-in an ideal transmission line- of the
 encoding-decoding scheme based on Fock states transmitted following a thermal distribution and photodetection at the stage of the receiver~\cite{Yuen}. We recall
  that the optimisation is performed under a constraint on the mean photons number in the information-bearing system. When it comes to realistic communications, however, the unavoidable presence of noise distorts the transmitted states and alternative encoding-decoding strategies may offer better performances. In fact, for a lossy bosonic channel, it has been shown that an amplitude coherent encoding (usually termed 
  "amplitude-based" scheme), where information is imprinted into the amplitude of coherent states and extracted via heterodyne or double-homodyne detection is indeed optimal. The capacity of the amplitude-based scheme achieved by heterodyne 
  detection is found to be
\begin{equation}\label{amplitudecapacity}
C_{\text{amp}}(\eta)=\log_{2}{(1+\eta\bar{n})},
\end{equation}
where $0<\eta<1$ denote the loss parameter and $\bar{n}$ stands for the mean number of photons of the coherent seed state.
\par
\begin{figure}[h!]
	\includegraphics[width=0.22\textwidth]{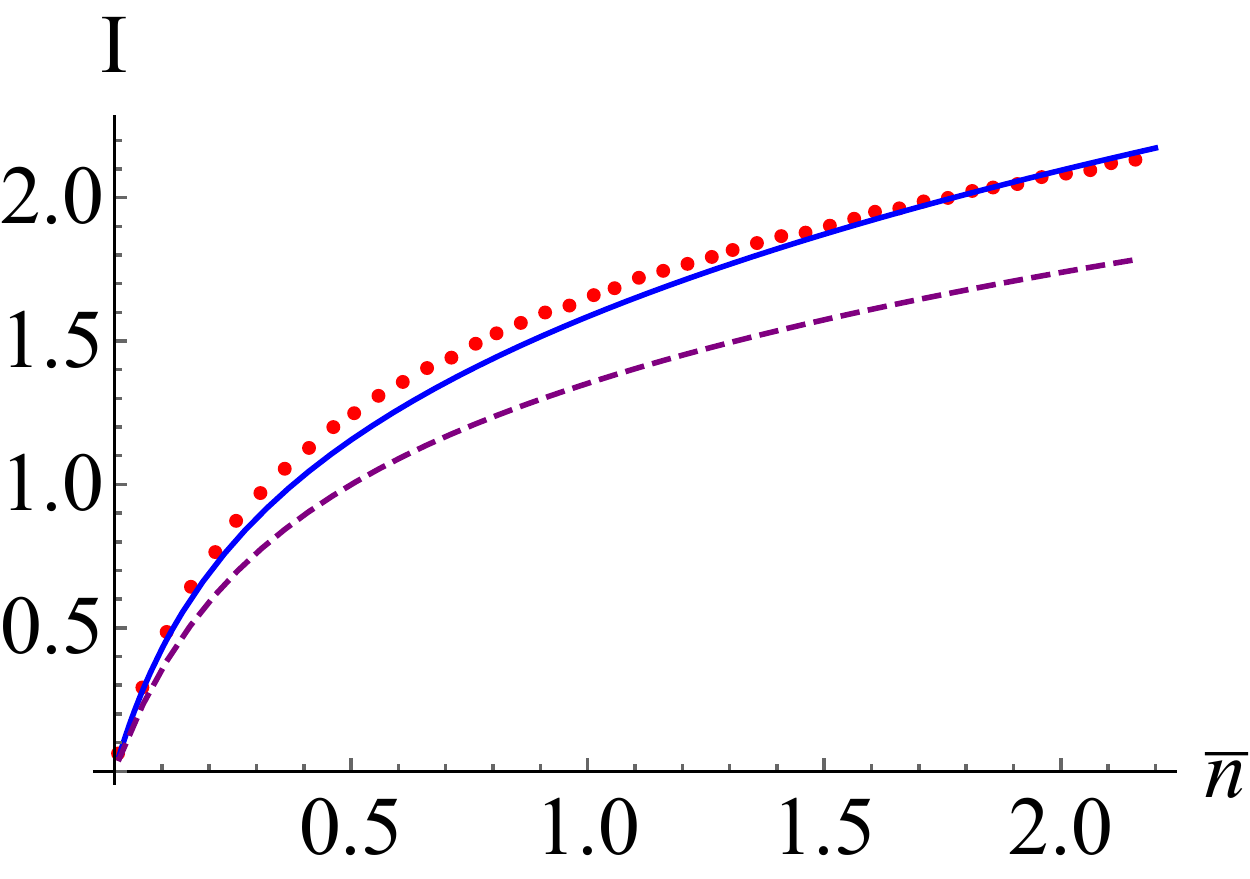}
	\includegraphics[width=0.25\textwidth]{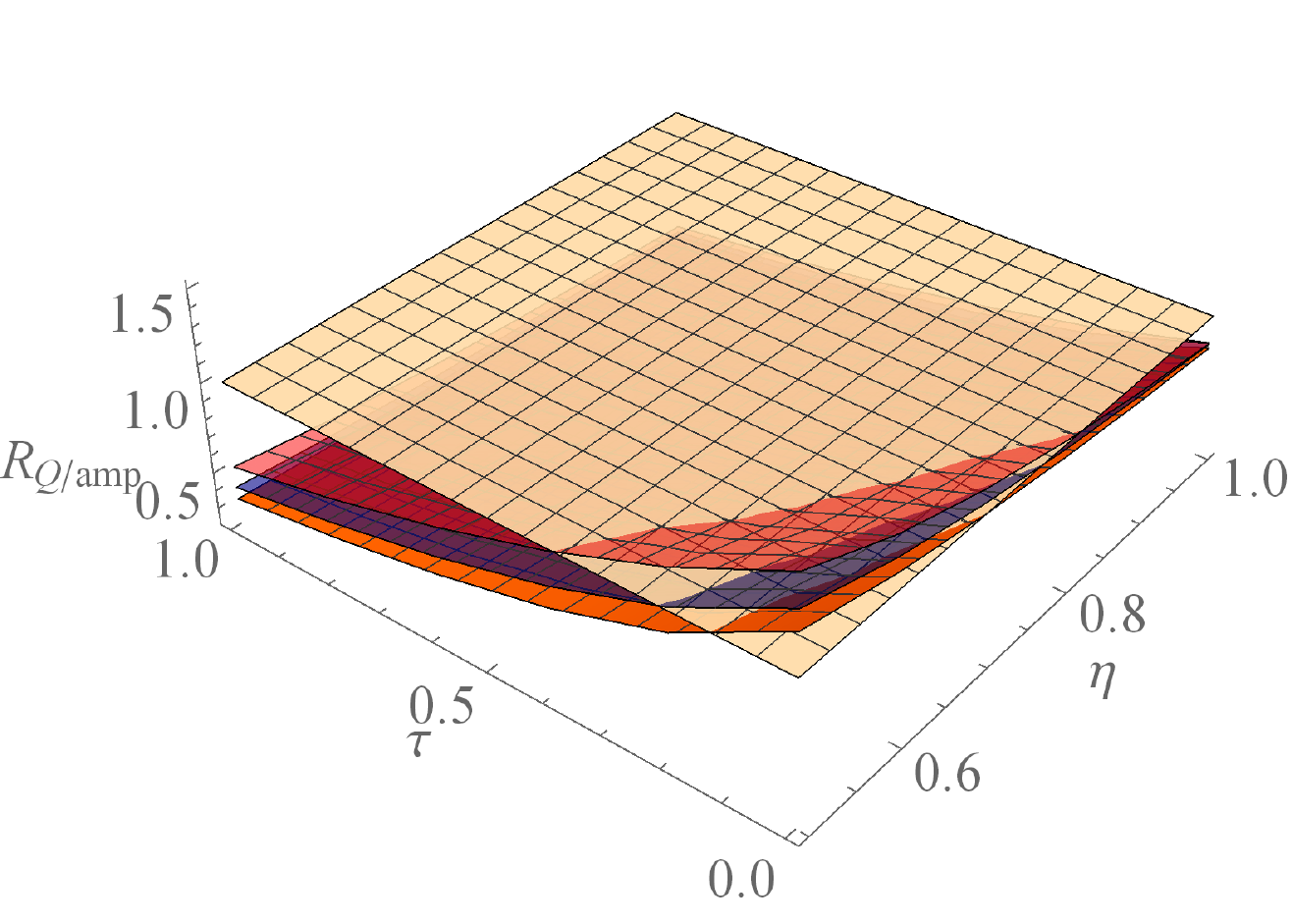}
	\caption{(Color online) Left panel: mutual informations $I_{\ID}$ (dotted red) and the  $I_{\QQ}$ (dotted purple) receivers together with the capacity of amplitude-based channel (solid blue) as functions of $\bar{n}$ for $M=30$. The considered quantities are evaluated for ideal channels, namely, $\eta=1$ and $\tau=0$. Right panel: 3D plots of the ratio $R_{\QQ/\text{amp}}=I_{\QQ}/C_{\text{amp}}$ as a function of $\eta$ and $\tau$ for different values of $\bar{n}$: from top to bottom $\bar{n}=0.2,1,2.$. The number of 
	symbols is set to  $M=20$.} \label{f:pskvsamp}
\end{figure}

In other to deepen our current analysis, we compare the performances of the 
PSK scheme assisted by the NLA to those of the amplitude-based one. First, 
we focus on the ideal channel, namely, $\tau=0$ and $\eta=1$. In the 
right panel of Fig~\ref{f:pskvsamp}, we show the mutual informations 
for both the ideal $I_{\ID}$ and the Q receivers $I_{Q}$ along with the 
capacity of the amplitude-based channel as functions of the mean photons number. We set the symbols ensemble size to $M=20$. Since we intend to establish a comparison between the two channels, the plots are realized for seed signals with equal energies. As it appears from the plots, the two considered schemes afford the same performances in the relevant regime of weak energies ($\bar{n}\ll 1$). However, in the remaining range of $\bar{n}$, the ideal receiver yields some trivial improvement when the average photons number don't exceed a certain threshold while the Q receiver becomes less efficient as $\bar{n}$ increases. 
\par
Let us now compare realistic transmission lines, i.e. phase channels with 
phase diffusion and amplitude channels with loss. Since the experimental 
implementation of canonical phase measurement remains unavailable, we will 
focus on the performances of the $Q$ receiver. In the right panel of Fig~\ref{f:pskvsamp}, we show a 3D plot of the ratio $R_{\QQ/\text{amp}}=I_{\QQ}/C_{\text{amp}}$ as a function of the noise parameters $\tau$ and $\eta$ for 
different values of the mean photons number ($\bar{n}=0.2,1,2$). 

The plot reveals the existence of a threshold value of $\bar{n}$, 
above which the phase channel provides better performances. Moreover, 
we notice that the region where the phase channel assisted by the 
NLA outperforms the amplitude-based one becomes larger as the 
average photons number decreases, thus proving its effectiveness 
in the relevant regime of weak signals.
\par
\subsection{Phase-shift keyed channels in the presence of dynamical noise}
The noisy channels considered so far were characterized by a static noise, i.e. a constant phase noise factor. As it happens, the dynamic of the states transferred along the transmission line is well described by a full quantum view under the Markov assumption. In various situations~\cite{Biercuk,Chin,Vasile,Schmidt,Thorwart}, the interaction of the quantum system of interest happens with a structured environment, thus the Markov approximation is no more appropriate. Generally, the theoretical description of quantum systems interacting with correlated in time surrounding degrees of freedom poses real issues. However, when the noise introduced by the environment present classical characteristics-as, for instance, Gaussian noise- the dynamics of the system may be properly modelled by a classical stochastic field (CSF)~\cite{Witzel,Trapani15}. Furthermore, beyond its simple formulation, the CSF description enables to investigate the impact of the memory effects induced by the environment on the system. 
\begin{figure}[h!]
	\includegraphics[width=0.2\textwidth]{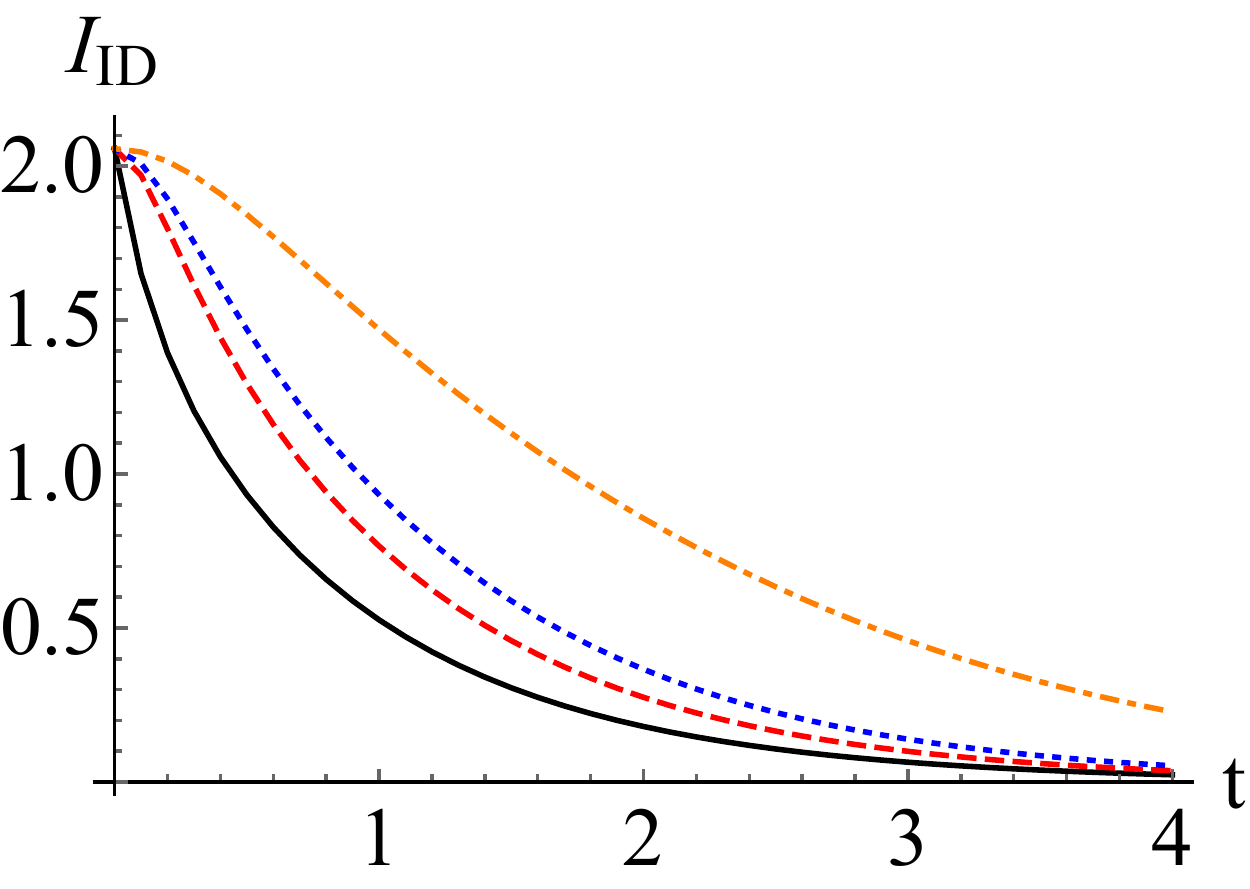}
	\includegraphics[width=0.22\textwidth]{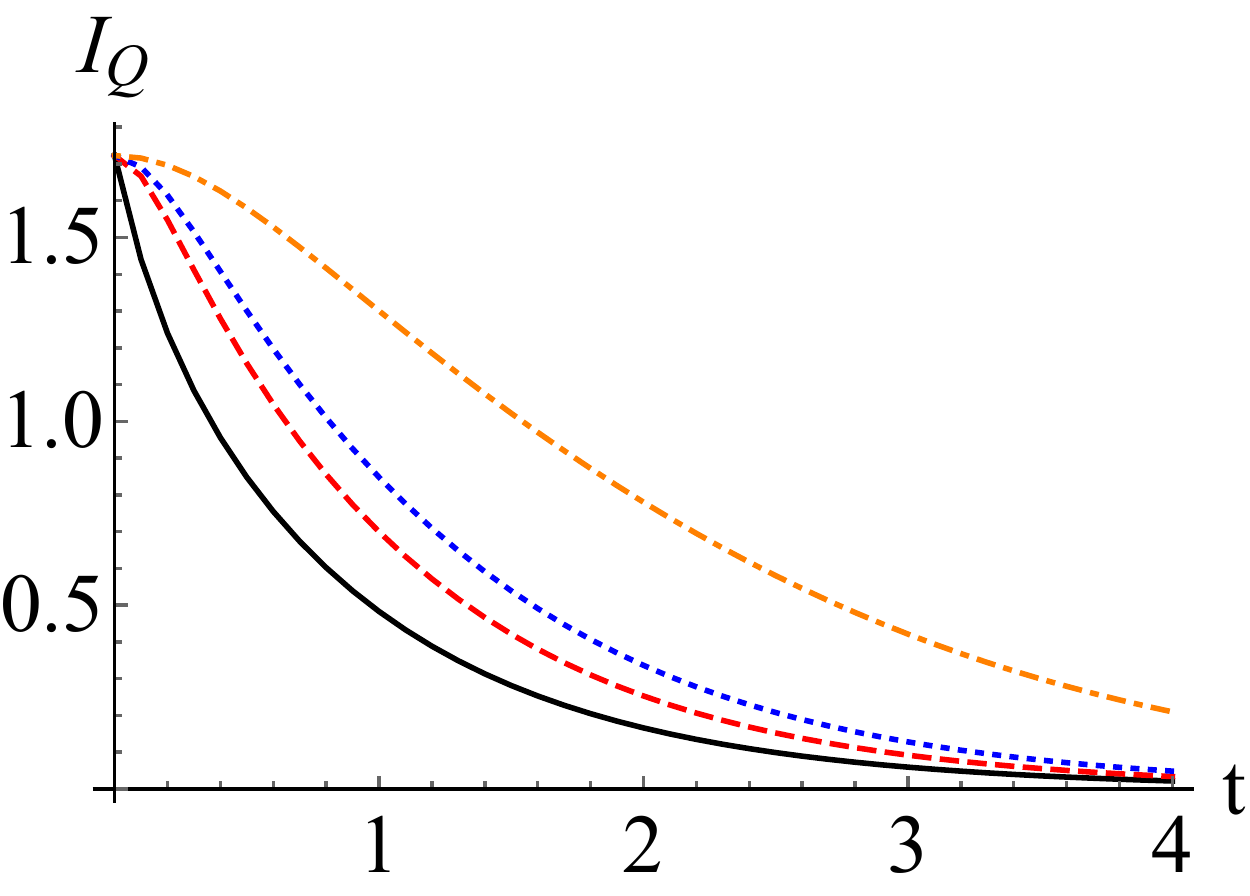}\\
	\includegraphics[width=0.22\textwidth]{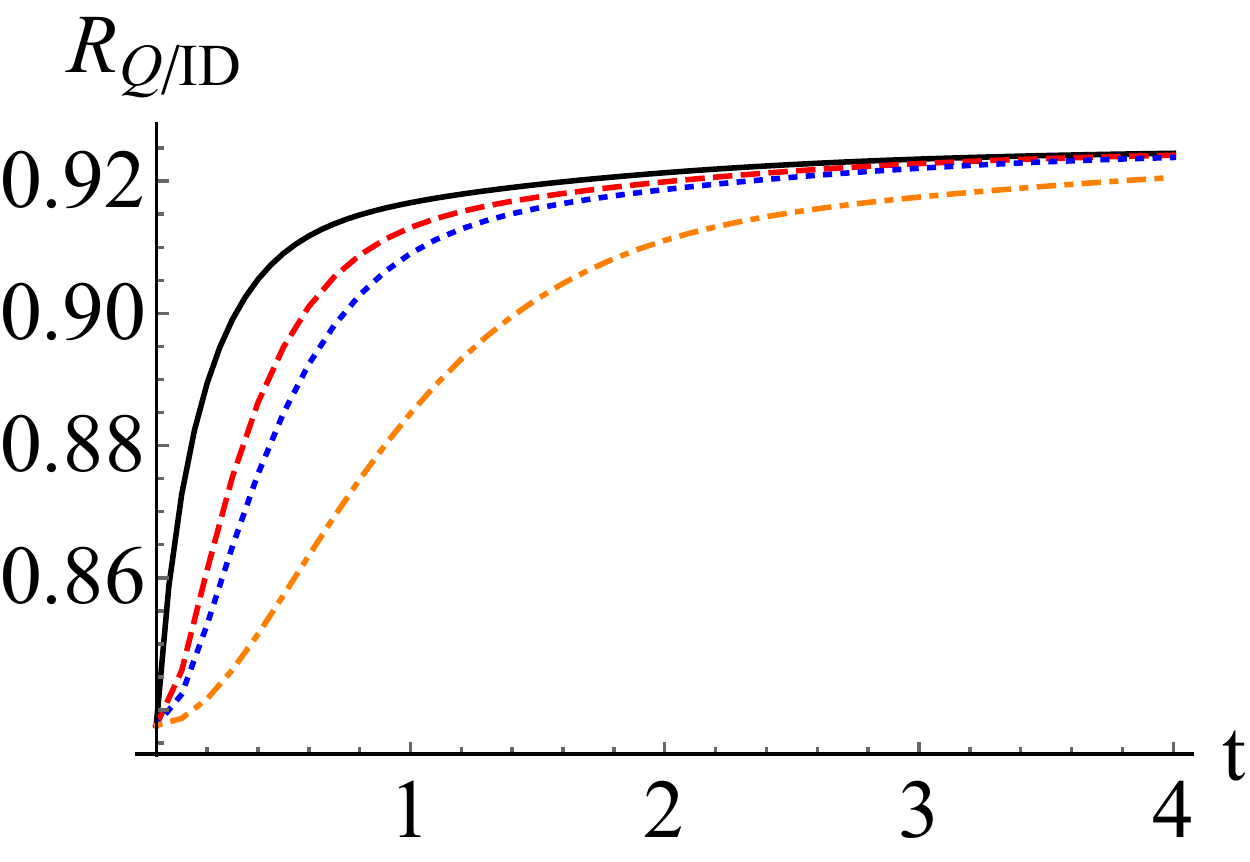}
	\includegraphics[width=0.25\textwidth]{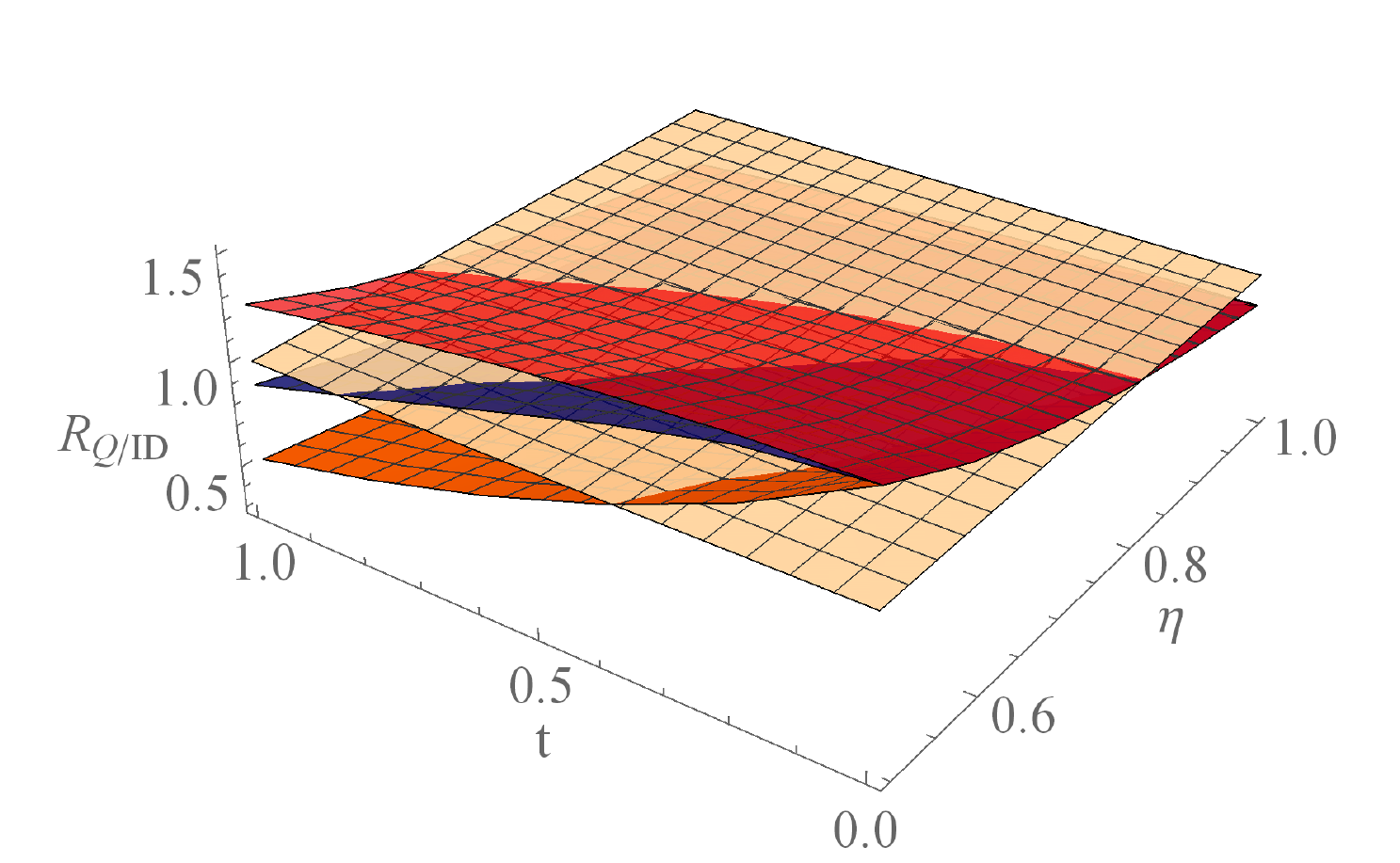}
	\caption{Color online) Upper panels: the mutual information $I_{\ID}$ (left) and $I_{\QQ}$ (right) as a function of the interaction time $t$ in the presence of 
	dynamical phase diffusion for different values of the characteristic correlation time $t_{E}$: $t_{E}=0.5$ (dashed red), $t_{E}=1$ (dotted blue),$t_{E}=5$ (dotdashed orange). We also report the plot of the mutual information in the presence of static noise (black solid). Lower left panel:  the ratio $R_{\ID/\QQ}=I_{\ID}/I_{\QQ}$ as a function of $t$ for the same  configurations of the upper panels. The colour 
	code is also kept unchanged. Lower right panel: 3D plot of the ratio 
	$R_{\QQ/\text{amp}}=I_{\QQ}/C_{\text{amp}}$ as a function of $\eta$ and 
	$t$ for different phase diffusion noises: from bottom to top, static 
	noise, noise with $t_{E}=1$ and noise with $t_{E}=5$. The remaining 
	parameters in all the plots are set as follows: $\bar{n}=2$, $M=20$, 
	$a=3$ and $\Gamma=1$ so as to be consistent with the static case. } \label{f:ratiodyn}
\end{figure}
\par
In the following, we will address phase communication channels in the presence 
of dynamical Gaussian noise. We adopt the physical model introduced in 
Sec~\ref{s:sec3}, where phase diffusion is modelled by a CSF. 
Given a seed state $\varrho(0)$, the evolved density matrix is given by Eq~(\ref{rhoevolved1}) and can be rewritten as:
\begin{equation}\label{rhoGaussian}
{\varrho}(t)=\int\! d\phi\, \mathcal{N}(\phi,\sigma)\,
{U}(\phi)\varrho(0){U}^{\dagger}(\phi)\,,
\end{equation}
where $\mathcal{N}(\phi,\sigma)$ is a normal distribution with zero mean and 
a time-dependant variance $\sigma(t)$. It entirely characterizes the stochastic 
Gaussian process (CSF) and is directly related to its autocorrelation function 
through Eq. (\ref{sigma}) \footnote{The Gaussian approximation for $\mathcal{N}(\phi,\sigma)$ is valid for $\sigma(t)\ll 2\pi$, otherwise a Von-Mises distribution for
the angular variable $\phi$ is used.}. The transformation~(\ref{rhoGaussian}) induced by the environment, takes an initial state $\varrho(0)$ to a statistical mixture of phase-shifted states distributed according to a Normal law around its initial phase. In the following, we will treat the case of the power-law (PL) process as an illustrative pattern of Gaussian CSF. The PL process is characterized by the kernel $K(t,t^{\prime})=\frac{a-1}{2}\frac{\gamma \Gamma}{\left( 1+\gamma \left|t-t^{\prime} \right|\right) ^{a} }$, where $\Gamma$ stands for the phase diffusion parameter and $\gamma=t_{E}^{-1}$, with $t_{E}$ being the correlation time of the environment. As it happens, for a null central frequency of the environment ($\omega=0$) the time-dependant variance $\sigma(t)$  reads
\begin{equation}\label{sigmat}
\sigma_{PL}(t)=\frac{\Gamma}{\gamma}\left[\frac{(1+\gamma t)^{2-a}+\gamma t(a-2)-1}{a-2} \right]. 
\end{equation}
In the limit where $\gamma \rightarrow \infty$, or equivalently, the environment's correlation time  is negligible with respect to the interaction time ($t_{E}\ll t$), the variance reduces to 
\begin{equation}\label{sigmamarkov}
\sigma_{PL}(t)\simeq\Gamma t,
\end{equation}
whereas for vanishing $\gamma$, namely, in the presence of consequential memory effects ($t_{E} \gg t$), it assumes the following form
\begin{equation}\label{sigmat2}
\sigma_{PL}(t)\simeq \frac{\Gamma t^{2}}{2}\left(a-1 \right) ,
\end{equation}
As it appears from Eq~(\ref{sigmamarkov}), where the variance is linear in 
time, we are back to the static situation where memory effects are negligible, 
and the CSF is well described in the Markovian approximation, thereby, the mutual information shows the same behaviour as in Fig~\ref{f:pskvsamp}.
\par
Let us now analyze the impact of memory effects on the information transferred along a phase diffusive channel. The mutual informations for both the ideal and the Q receiver are evaluated by performing the replacement $e^{-\frac{1}{2}d^{2}\tau}\rightarrow e^{-\frac{1}{2}d^{2}\sigma(t)}$ in the expressions (\ref{qsI}) and (\ref{qsQ}) respectively. In the upper panels of Fig~\ref{f:ratiodyn}, we report variations of the mutual informations $I_{\ID}$ and $I_{\QQ}$ with respect to the interaction time $t$ for different values of the environment's correlation time $t_{E}$. For the sake of comparison, we also show the mutual information for the static phase noise. The plots reveal better performances for both the ideal and Q receiver in the presence of dynamical noise. Moreover, we observe that the larger is the correlation time, better preserved is the mutual information. In fact, the decay rate drops for consequential memory effects whereas being at its maximum for vanishing $t_{E}$ as illustrated by the transition (Eq(\ref{sigmamarkov})$\rightarrow$ (Eq(\ref{sigmat2}))) from linear to quadratic variations in time of $\sigma (t)$. In the lower panel, right side, we show the ratio $R_{\QQ/\ID}=I_{\QQ}/I_{\ID} $ as a function of the interaction time $t$ for different values of $t_{E}$. We observe a decreasing ratio with respect to the correlation time, denoting that the dynamical phase noise disadvantages the Q receiver compared with the ideal one. 
\par
Let us now compare the performances to those of an amplitude-based channel 
in the presence of noise. Once again, we consider a dynamical noise originated 
from a long range power-Law process. Our results are depicted in the lower 
right panel of Fig~\ref{f:ratiodyn}, where we show a 3D plot of the ratio 
$R_{\QQ/\text{amp}}=I_{\QQ}/C_{\text{amp}}$ as a function of the amplitude 
loss parameter $\eta$ and the interaction time $t$, for two values of the
 correlation time of the environment ($t_{E}=1,5$). In order to 
 emphasize the contribution coming from the memory effects, we also report 
 the 3D plot of the ratio in the presence of static noise. It is worth noting 
 that we have considered here the regime of weak energies ($\bar{n}=0.2$). We 
 clearly notice that the environment memory effects enhance the performances 
 of the noisy phase-based channel assisted by the NLA with respect to the 
 amplitude-based one. This is illustrated by the increased area where the 
 phase channel outperforms its amplitude counterpart for higher time-correlated environments. Overall, we conclude that memory effects are a resource to 
 preserve information in the presence of phase diffusion.
\section{Conclusion}\label{s:sec6} 
We have investigated quantum phase communication channels assisted by 
probabilistic noiseless linear amplification and assessed their performances 
in presence of static and dynamical phase noise. 
\par
At first, we have shown that in the presence of Markovian noise, NL-amplification
of the coherent seed signal improves the performances for both ideal and feasible
receivers. Moreover, upon comparison with lossy coherent states amplitude-based 
scheme we have shown the existence of a threshold on the loss and phase noise parameters, above which phase channels better preserve the transfer of 
information. Then, we have shown that in the presence of time-correlated noise, 
leading to dynamical non-Markovian phase diffusion, the interplay between the 
use of NLA and the memory effects provide a noticeable improvement of 
performances, i.e memory effects better preserve the information transferred 
along the transmission line. 
\par
Overall, our results prove that quantum phase communication channels may be
of interest for applications with current technology, and pave the way 
for their implementations in realistic scenarios.
\section*{Acknowledgement}
This work has been supported by the Center for Cyber-Physical 
Systems (Khalifa University) through the Grant Number 
8474000137-RC1-C2PS-T3 and by the Foundation for Polish Science
through the TEAM project {\em Quantum Optical Communication Systems}.
The authors thank Abdelhakim Gharbi and Ernesto Damiani for 
useful discussions. 

\end{document}